\documentclass[aps,letterpaper,amssymb,twocolumn,pra,showpacs,floatfix]{revtex4}
\usepackage{amssymb}
\usepackage[usenames, dvips]{color}

\usepackage{graphicx}
\usepackage{amsmath}
\usepackage{times}
\usepackage{subfigure}


\begin{document}

\title{Two-Dimensional Electron Gas with Cold Atoms in Non-Abelian Gauge Potentials}

\author{Indubala I. Satija$^{1,2}$, Daniel C. Dakin$^{3,1}$, J. Y. Vaishnav$^{2}$ and Charles W. Clark$^{2}$}
\affiliation{$^{1}$ Dept. of Physics, George Mason U., Fairfax, VA 22030, USA}
\affiliation{ $^{2}$ Joint Quantum Institute, National Institute of Standards and Technology,
Gaithersburg, MD 20899, USA}
\affiliation{ $^{3}$ Optical Air Data Systems, 10781 James Payne Court, Manassas, VA 20110 }

\date{\today}
\begin{abstract}
Motivated by the possibility of creating non-Abelian fields using cold atoms in optical lattices, 
we explore the richness and complexity of non-interacting two-dimensional electron gases
(2DEGs) in a lattice, subjected to such fields.  In the continuum limit, 
a non-Abelian system characterized by a two-component ``magnetic flux" describes a
harmonic oscillator existing in two different charge states (mimicking a particle-hole pair) where the coupling between the states is determined by the non-Abelian parameter, namely the difference between the two components of the ``magnetic flux".
A key feature of the non-Abelian system is a splitting of the Landau energy
 levels, which broaden into bands, as the spectrum depends explicitly on the transverse momentum.  
These Landau bands result in a coarse-grained ``moth", a continuum version of the generalized Hofstadter butterfly.  Furthermore, the bands overlap, leading to effective relativistic effects.  Importantly, similar features also characterize the corresponding two-dimensional lattice problem when at least one of the components of the magnetic flux is an irrational number. The lattice system with two competing ``magnetic fluxes" penetrating the unit cell provides a rich environment in which to study localization phenomena. Some unique aspects of the transport properties of the
non-Abelian system are the possibility of inducing localization by varying the quasimomentum, and the absence of localization of certain
zero-energy states exhibiting a linear energy-momentum relation.  Furthermore, non-Abelian systems provide an interesting localization
scenario where the localization transition is accompanied by a transition from relativistic to non-relativistic theory.

\end{abstract}
\maketitle

\section {Introduction }

Methods for creating fields that couple to neutral atoms in the same way 
that electromagnetic fields couple to charged particles have created the exciting possibility of studying the effects
of a generalized magnetism using cold 
atoms.\cite{mono,NJP,NA}   Using laser induced hopping, 
a controlled phase can be imposed upon particles
moving along a closed loop in an optical lattice.
The associated synthetic fields can be sufficiently strong to enter the regime of exotic magnetic phenomena that have
been difficult to explore in condensed matter experiments,
 such as
the fragmented fractal spectrum of a two dimensional electron gas (2DEG) in a magnetic
field, the famous ``Hofstadter butterfly"\cite{Hofs}. 
Such fields need not obey Maxwell's equations, thus 
providing the possibility of discovering fundamentally new physics.
\cite{Escher}. For example, we discuss here the generation of non-Abelian 
fields,
 by using cold atoms that occupy two Zeeman states in the hyperfine ground 
level~\cite{NA}; these two states may be thought of as ``colors" of the gauge fields, 
and such a system may be used to simulate lattice gauge theories in (2+1) dimensions. 
Other potential applications of non-Abelian fields are the creation of counterparts of
magnetic monopoles~\cite{mono},  and 
integer and fractional quantum Hall effects.\cite{chern} 

In this paper, we adopt the 2DEG as a motif for the study of cold atom systems.
The subject of 2DEGs in a magnetic field is a textbook topic~\cite{Prange}, as the problem maps
 to a one-dimensional harmonic oscillator. The discrete energy levels of the oscillator are the 
Landau levels that describe free particle energies in terms of the quantized units 
$\hbar \omega_c$, where $\omega_c=eB/mc$ is the cyclotron frequency of the 
corresponding classical motion. Each level is highly degenerate,
reflecting the fact that a classical electron  
spirals about a line parallel to the magnetic field,
with an arbitrary center in the transverse plane. The degree of degeneracy is equal to
$L^2/2\pi \delta^2$ where $\delta=\sqrt{\hbar c/eB}$ is the magnetic length and $L^2$ is the area of the system.

Beginning with the celebrated work of Onsager~\cite{Ons}, Harper~\cite{Harper} and 
then Hofstadter~\cite{Hofs}, the subject of 2DEGs in a crystalline lattice in
a magnetic field has 
fascinated physicists as well as mathematicians. 
In the presence of a lattice, each Landau energy level splits into $Q$ bands, where the rational 
number $P/Q$ is the magnetic flux through the unit cell in units of the magnetic flux quantum (fluxoid).
The heart of the problem is the two competing periodicities 
related to the ratio of the reciprocal of the cyclotron frequency and the period of the motion of the electron
in the periodic lattice.
Two key aspects that have been explored extensively are the 
exotic multifractal spectrum (Hofstadter butterfly), and the metal-insulator transition obtained by tuning the ratio of the tunneling along the two directions of the lattice~\cite{Harper}. 
Recent studies have shown that these properties can be demonstrated
using ultracold atoms in an artificial magnetic field ~\cite{NJP, Drese}.
This paper revisits the metal-insulator transition when the 2DEG is 
subjected to a non-Abelian gauge field which is a natural generalization of the uniform magnetic field.
Such fields yield a much richer spectral and transport landscape than is encountered in the Abelian case.

The generic experimental setup for producing non-Abelian $U(2)$ gauge fields that we consider here,
consists of a
two-dimensional optical lattice populated with cold atoms that occupy two hyperfine states~\cite{NJP,NA}.
Such systems exhibit three competing length scales , associated with two distinct ``magnetic fluxes" 
(denoted by $\alpha_1$ and $\alpha_2$) that penetrate the unit cell. 
Our aim is to describe some of the generic properties
of such systems. Although our main focus is on optical lattices,
we first discuss the corresponding continuum problem, where the
the infinite degeneracy of the Landau levels is lifted by non-Abelian interactions.
%The continuum system maps onto the problem of two oppositely-charged coupled harmonic oscillators,
%with a coupling constant proportional to the non-Abelian
%parameter $(\alpha_1-\alpha_2)$.
The continuum problem mirrors 
some of the features subsequently encountered in the lattice system.

In the discussion of the metal-insulator transition in the lattice, we focus on the ground state as well as the
states at the band center. These two cases are respectively relevant for experimental systems involving 
Bose condensates and fermionic gases near half-filling.
Some of these results were described in an earlier paper~\cite{prl}.
In addition to a detailed analysis, here we describe new results,
such as the simulation of relativistic phenomena using cold atoms in non-Abelian fields.
By tuning lattice anisotropy, we can implement relativistic as well as non-relativistic dynamics,
 with a particular focus on the effects of disorder.
Simulation and detection of Dirac fermions using single-component cold atoms 
in a hexagonal lattice was recently proposed~\cite{duan}.
The systems we propose here provide
the possibility of observing relativistic particles  and also of studying their localization properties.
We show that the non-Abelian systems provide an experimental realization of the defiance of localization by disordered
relativistic fermions, a topic that has been the subject of extensive study~\cite{Fradkin}.

In Section II, we introduce non-Abelian gauge fields and the corresponding effective ``magnetic fields."
Section III examines the continuum limit of a single particle in a non-Abelian gauge field. 
In Section IV, we discuss lattice systems subjected to these fields, and describe methods of calculation.
In Section V, we study various spectral characteristics of the non-Abelian lattice systems. 
There, following long establised practice for studying metal-insulator transition in
Abelian systems, we fix $\alpha_1=(\sqrt{5}-1)/2$,
the golden mean, which we denote as $\gamma$. The irrationality of $\alpha_1$
ensures the existence of a localization transition.\cite{Harper, Sok}
For $\alpha_2$,  we consider a selected set of
both rational and irrational values.
%For rational case, we confine ourselves to simple fractions such as $1/2, 1/3, 1/4$, etc. while for
%irrational case, we choose $\alpha_2=\gamma^3, \gamma^4$, which are harmonics
%of $\gamma$ since $\gamma^3=2\gamma-1$ and $\gamma^4=2-3\gamma$.
Sections VI and VII discuss localization properties of the states at the band center ($E=0$) and at 
the band edge.  The localization of the $E=0$ states brings out some of the most important 
features of the non-Abelian cases, including the dependence of the transition upon 
a conserved momentum.
Furthermore, a unique aspect of the the non-Abelian system, namely the defiance of localization
of the $E=0$ states, 
emerges when the energy-momentum relation mimics the behavior of relativistic particles. 
Section VIII describes the experimental realization of the metal-insulator transition in cold atom lattices. 
%In the Appendix, we briefly discuss the Aharonov-Bohm and Aharonov-Casher effects.

\section{ Non-Abelian Gauge Fields }

Effective non-Abelian vector potentials arise naturally in systems where the atoms have $N$ degenerate internal states.  The most general vector potential couples the states, and thus gives rise to a $U(N)$ gauge symmetry. We here consider the case where $N=2$.  In our treatment of the non-Abelian case, we follow the convention of an earlier study~\cite{NA}, adopting its form of vector potential,
\begin{eqnarray} 
\vec{A}&=&\frac{\hbar c}{ea}\left(\frac{\pi}{2}\left( \begin{array}{cc} -1 & 1\\ 1& -1\\ \end{array}\right), \:
2\pi \frac{x}{a} \left(\begin{array}{cc} \alpha_1 & 0\\ 0 & \alpha_2\\ \end{array} \right ), \: 0 \right ).
\label{eq:contvp}
\end{eqnarray} 
The $\alpha_i$ determine the ``magnetic fluxes" of the lattice with lattice constant $a$.

Equation~(\ref{eq:contvp}) is in the Landau gauge: $\vec{A}(x,y) = (A_x,A_y(x),0)$, where $A_x$ is a constant
and $A_y$ depends only on $x$. 

We rewrite the vector potential in terms of Pauli matrices 
$\sigma_i$, separating the Abelian and the non-Abelian parts of the gauge field, \begin{eqnarray}
\vec{A}&=&\frac{\hbar c}{ea}\left[-\frac{\pi}{2}(I-\sigma_x) \hat{\mathbf{x}}+ 2\pi \frac{x}{a} \alpha I \hat{\mathbf{y}} + \Delta \, \frac{x}{a}
\sigma_z  \hat{\mathbf{y}} \right],
\label{vectorpotential}
\end{eqnarray}
where we have defined quantities $\alpha= (\alpha_1+\alpha_2)/2 $ and $\Delta = \pi (\alpha_1-\alpha_2)$.
Here $\sigma_a$, $a=x,y,z$ denotes Pauli matrices. The parameter $\Delta$ characterizes the non-Abelian feature of the system.

For non-Abelian fields, the effective ``magnetic field"
is given by,
\begin{equation}
\hat{\mathbf{B}}=\nabla\times\hat{\mathbf{A}}-\frac{ i e}{\hbar c }\hat{\mathbf{A}}\times\hat{\mathbf{A}}.
\label{eq:yangmills}
\end{equation}
The origin of the extra term $\hat{\mathbf{A}}\times\hat{\mathbf{A}}$ can be traced to the 
commutator for the generalized velocity operator $(p-\frac{e}{c}A)/M$,
\begin{eqnarray*}
[v_n, v_m] &=&\frac{\hbar e}{M^2c}\left(\partial_n A_m-\partial_m A_n -\frac{ie}{\hbar c}[ A_n, A_m]\right)\\
&=& \frac{i\hbar e}{M^2c} \epsilon_{mnr}B_r.
\end{eqnarray*}  
For the vector potential in Eq. (\ref{eq:contvp}), this gives
\begin{equation}
\hat{B}_z = \hat{B}_0 + \Delta \left(\frac{\hbar c}{ea^2}\right)(\hat{\mathbf{\sigma}}_z -\pi \frac{x}{a}
\hat{\mathbf{\sigma}}_y ),
\end{equation}
where $B_0= 2\pi\alpha (\frac{\hbar c}{ea^2})$. Thus, $\alpha=B_0 a^2/( 2\pi\hbar c/e)$ describes the Abelian flux quanta
penetrating per unit cell of the lattice.
The non-Abelian gauge potential generates a non-uniform magnetic field, as $\hat{\mathbf{B}}$ depends explicitly on the spatial coordinate $x$ when $\Delta \ne 0$.

%\begin{equation}
%\hat{\mathbf{B}} = \left[ B_0 + \Delta ( \hat{\sigma}_z- x \pi\hat{\sigma}_y ) \pi \right] \hat{\mathbf{z}}. 
%\end{equation}
%It should be noted that
%although $\hat{\mathbf{B}}$ diverges, the magnetic flux penetrating the unit cell is
%finite.

\section { Continuum Limit of the Non-Abelian System } 

We now consider the continuum limit of the non-Abelian problem. 
Although the $\hat{\mathbf{A}}$ of the Eq. (\ref{vectorpotential}) is ill-defined  in the continuum limit
$a \rightarrow 0$, the study is useful in illustrating some key aspects of the non-Abelian systems.
%In our analysis, we will set $\hbar c/ea$ to be unity.
In general, continuum problems can also be experimentally realized, as in Ref.~\cite{mono}.

It can be shown, after some algebra, that the two-component continuum Hamiltonian 
$\hat{\mathbf{H}}_c  = (\hat{\mathbf{p}}-\frac{e}{c}\hat{\mathbf{A}})^{2}/(2M)$ resulting from the vector potential 
in Eq. (\ref{vectorpotential}) is gauge-equivalent to

\begin{eqnarray}
\hat{\mathbf{H}} & = &\frac{1}{2M}\left(\begin{array}{cc}
%(\hat{p}_{x}+\beta)^{2}+M^2\omega^2 x^2 & C\Delta (x-x_0)^2 \\
(\hat{p}_{x}+\beta)^{2}+V(x) & C\Delta (x-x_0)^2 \\
C\Delta (x-x_0)^2 & (\hat{p}_{x}-\beta)^{2}+V(x)
\end{array}\right)
\label{nah}
\end{eqnarray}

up to a $k_y$ dependent term. The transverse momentum $k_y$ is a conserved quantity as the Hamiltonian
$\hat{\mathbf{H}}_c$ with $\vec{A}$ given by Eq ~\ref{vectorpotential} is cyclic in $y$.
Here, $\beta=\frac{\hbar \pi}{2 a}$,  $C=8\pi^2 \alpha (\beta/a)^2$
,$V(x)= M^2 \omega^{2} x^{2}/2$ 
with $\omega=\frac{2 \pi \beta}{ma} \sqrt{(\alpha_{1}^2 + \alpha_2^2)/2}$ 
and $x_0 = 2 \pi a  \frac{\hbar k_y \beta} {M^2 \omega^2 a} \frac{ \alpha_1 \alpha_2}{ \alpha}$.
%    x_0 = 2 \pi a  \frac{\hbar k_y}{\beta} \frac{ \alpha_1 \alpha_2}{ (\omega/\frac{2 \pi \beta}{ma})^2 \alpha} 
This particular form of the Hamiltonian provides a new, illuminating picture of the non-Abelian problem; 
the particle behaves as a two-component harmonic oscillator existing in a positive as well 
as a negative charge state.  The physics of this system is that of a 
particle-hole pair, with the non-Abelian parameter $\Delta$
, governing the coupling between states.

\begin{center}
\begin{figure}[h]
\begin{centering}
\includegraphics[width=0.5\textwidth,keepaspectratio]{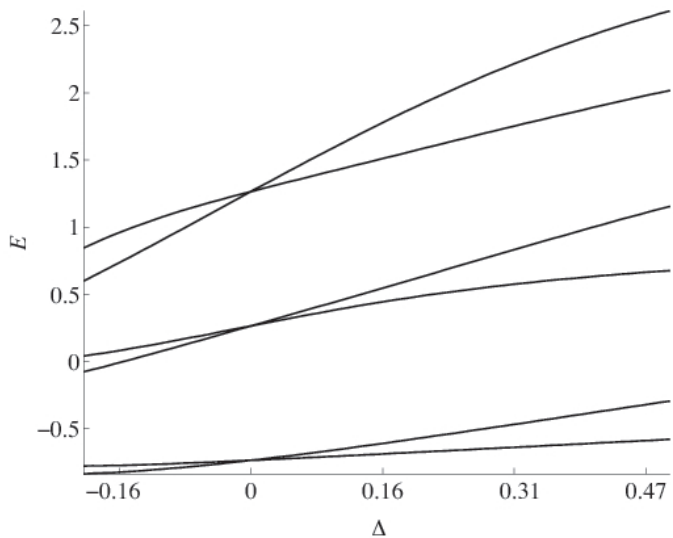}
\end{centering}
\caption{Six lowest energy levels of the continuum non-Abelian Hamiltonian
in Eq.~(\ref{nah}), $\beta=\pi/2,M=1,k_y=1$ and $2\pi\alpha_{1}=1$.
The levels are equally spaced in (and only in) the Abelian case, $\Delta = 0$, where
the system reduces to two decoupled harmonic oscillators.}
\label{fig:projection}
\end{figure}
\end{center}

The spectrum of $\hat{\mathbf{H}}_c$ is obtained by numerical diagonalization in a basis of
harmonic oscillator wave functions with frequency $\omega$.
Figure~(\ref{fig:projection}) shows the six lowest energy levels. For fixed $k_y$, 
at $\Delta=0$, each Landau level is two-fold degenerate.
For $\Delta \ne 0$ the degeneracy is lifted and the eigenstates become
entangled states of a particle-hole pair.

\begin{center}
\begin{figure}
\begin{centering}\includegraphics[width=0.5\textwidth,keepaspectratio]{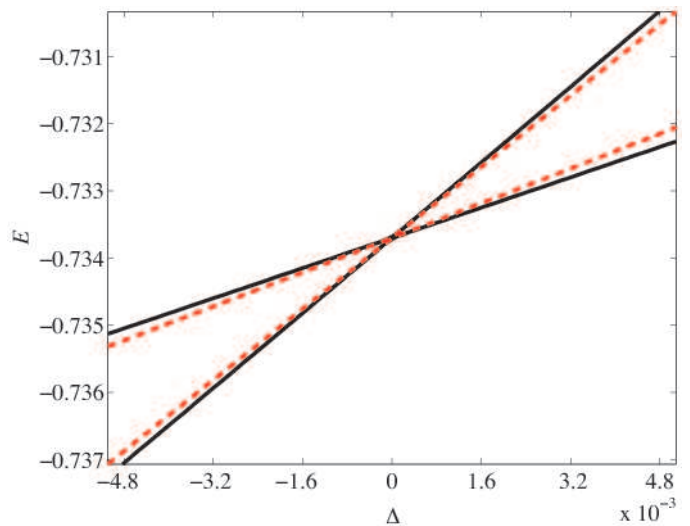}
\par\end{centering}
\caption{(color online) The ground state splitting obtained numerically 
(see Fig.~\ref{fig:projection}) (solid black), compared with that 
obtained by perturbation theory (dashed red) about the Abelian point $\Delta=0$.} \label{fig:perturb}
\end{figure}
\end{center}

\begin{figure}[htbp]\includegraphics[width=\columnwidth]{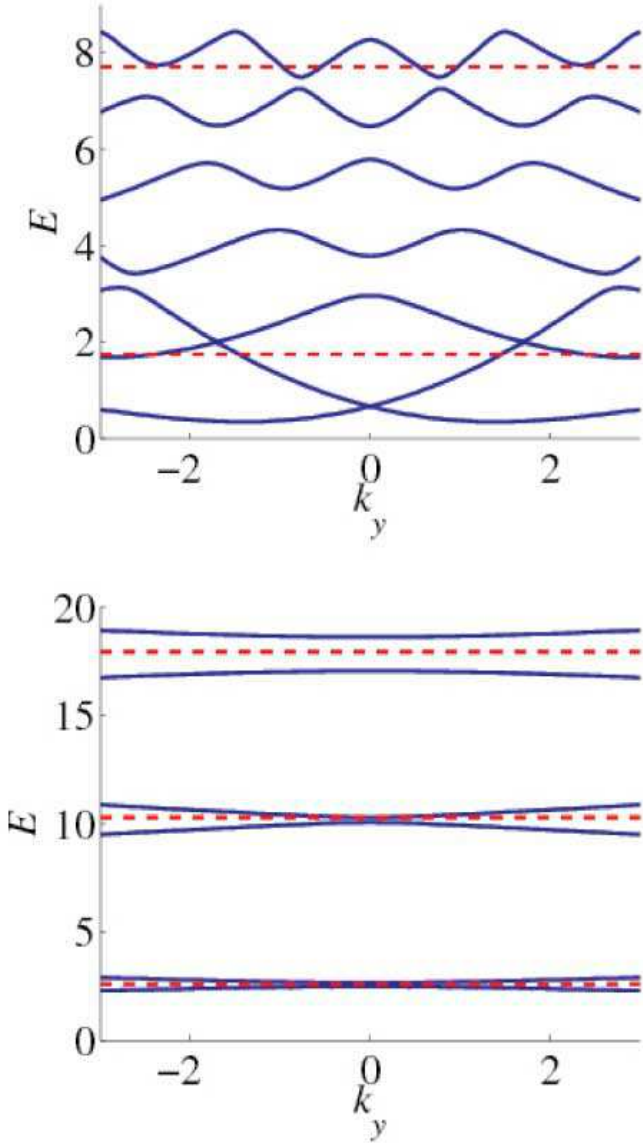}
\leavevmode \caption{(color online) The three lowest energy levels of the non-Abelian system with
$\alpha_1 = 1.32$ and $\alpha_2 = 0.253$ (top), $\alpha_2 = 1.11$ (bottom).
The dashed line shows the Abelian levels corresponding to $\sqrt{\frac{1}{2}(\alpha_1^2 + \alpha_2^2)}$.
}\label{fig:Econ}\end{figure}

As Fig.~\ref{fig:projection} shows, the energy levels are equally spaced only for the Abelian case, 
$\alpha_1=\alpha_2$. We can explicitly understand the splitting of each Landau level 
via degenerate perturbation theory, with $\Delta$ as a small parameter and 
using degenerate eigenstates, 
\begin{eqnarray*}
f_{1}^n= e^{i k_y y}\left( \begin{array}{cc} e^{-i \beta x} \psi_n(x)\\0 \end{array}\right), \:
f_{2}^n=e^{ik_y y}\left(\begin{array}{cc} 0\\ e^{i \beta x} \psi_n(x)  \end{array} \right ),
\end{eqnarray*}
where $\psi_n(x)$ are the eigenstates of the corresponding harmonic oscillator.
Figure \ref{fig:perturb} compares the perturbative splitting of the lowest Landau 
level with the numerical result.
%For the ground state, the first order correction can be written as,
%\begin{equation}
%\delta E_0= \pm  \Delta\sum_{n=0}^{n=4}{c_n k_y^n}, \label{eq:perturbation}
%\end{equation}
%where the $c_n$ are constants that depend upon the $\alpha_i$.
%Eq. (\ref{eq:perturbation}) illustrates the splitting of the levels, and the explicit dependence on $k_y$ away from the Abelian case.  Fig.~\ref{fig:perturb} shows that the perturbative calculations are in excellent agreement with the numerical results for small $\Delta$.

Figure~\ref{fig:Econ} shows the variation of the energies with $k_y$, obtained numerically.  
In highly non-Abelian cases, the energies bear no relation to their Abelian values. 
%and the bending of the energies in $k_y$ results in crossings and avoided crossings (top) 
%yielding a linear dispersion relation, and concomitant relativistic effects.  
Close to the Abelian limit (bottom) the energy levels are simply split around the Abelian energies.
The energies oscillate with $k_y$, resulting in actual and avoided crossings (i.e., the Landau bands overlap). 
In the vicinity of the crossings, the bands exhibit a linear dispersion relation.
As we shall discuss, these features reappear in the corresponding problem of the non-Abelian gauge field on an optical lattice.

\begin{figure}[htbp]\includegraphics[width=\columnwidth]{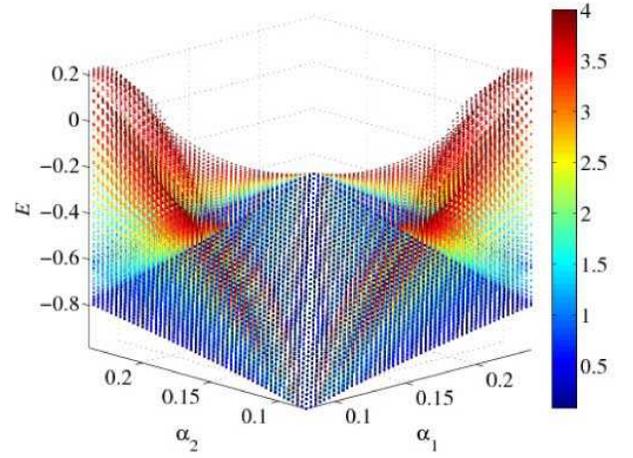}
\leavevmode \caption{(color online) The continuum version of the Hofstadter ``moth"~\cite{NA}. 
This plot shows the energies as a function of $\alpha_1$ and $\alpha_2$ for a range of $k_y$; 
the color scale indicates the range of $k_y$.  Along the line $\alpha_1=\alpha_2$ , 
the Abelian ``backbone" of the moth, there is no $k_y$ dependence.}
\label{cmoth}\end{figure}

Figure~\ref{cmoth} summarizes the effects of the non-Abelian gauge potential on the lowest Landau level of the corresponding
Abelian problem. The figure describes the continuum limit of the Hofstadter ``moth"~\cite{NA}, 
which is the generalization of the Hofstadter ``butterfly" as the underlying gauge field becomes non-Abelian. 
This coarse-grained ``moth" illustrates the symmetry breaking feature of the non-Abelian system as it
lifts the degeneracy of the corresponding Abelian problem.

\section {Two-Dimensional Lattice in Non-Abelian Gauge Fields }

Our starting point is a tight binding model (TBM) of a particle moving on a two-dimensional rectangular lattice $(x,y)$, with lattice constants $(a,b)$
and nearest-neighbor hopping characterized by the tunneling amplitudes $(J, J\Lambda)$.
When a weak external vector potential,  
$\vec{A}(x,y) = (A_x,A_y,0)$, is applied to the system, the Hamiltonian,
\begin{equation*}
\hat{H}= -J\left[\cos\left((p_x-\frac{e}{c}A_x)\frac{a}{\hbar}\right)+ \Lambda \cos\left((p_y-\frac{e}{c}A_y) \frac{b}{\hbar}\right)\right],
\end{equation*}
where $\vec{p}$ is the momentum operator.
Alternatively, the Hamilonian of a 2DEG on a lattice in the presence of a magnetic fied can be written as,
\begin{equation}
H=-\sum_{<ij>} J_{ij} c_j^{\dagger} c_i e^{i\theta_{ij}} + H.C.
\end{equation}
where $c_i$ is the usual fermion operator at site $i$. The $J_{ij}$ is the nearest-neighbor anisotropic hopping
with values $J$ and $J\Lambda$ along the $x$ and the $y$-direction. 

The phase factor $\theta_{ij}=-\theta_{ji}$ defined on a link $<i,j>$
is identified as $(2\pi e/ch) \oint {\bf A} . {\bf dl}$, where ${\bf A}$ is the vector potential, and
\begin{equation}
frac{1}{2\pi} \sum_{unit {} cell} \theta_{ij}=e/ch \oint {\bf A} . {\bf dl}= \frac{1}{\Phi_0} \oint {\bf B} . {\bf dS}
\end{equation}
is the magnetic flux penetrating the unit cell in units of magnetic flux quantum, $\Phi_0=ch/e$.

We denote the eigenfunction (projected onto the $x,y$ basis),
corresponding to the eigenvalue equation $H |\Psi> = E | \psi>$ 
as $\Psi(x,y)$.
With the transverse wave number of the plane wave
as $\tilde{k}_y=k_y/a$, the wave function can be written as: $\Psi(ma,na)=e^{i 2\pi k_y n} {\bf g}_m$ with $x=ma$ and $y=na$.
%where $\alpha=Bab/(hc/e)$ and $E(\alpha)$ is the energy in units of $-J$.

Subsituting the vector potential defined in the Eq. ~\ref{vectorpotential}, the two-component vector
${\bf g}_m=\left (\begin{array}{cc} \theta_m\\ \eta_m \end{array} \right )$ can be shown to result in the following equations,
\begin{eqnarray*}
\left (\begin{array}{cc} \theta_{m+1}\\ \eta_{m+1} \end{array} \right )+
\left (\begin{array}{cc} \theta_{m-1}\\ \eta_{m-1} \end{array} \right )-
\left (\begin{array}{cc} 0 &E-V_m\\ E-U_m & 0 \end{array} \right )
\left (\begin{array}{cc} \theta_m\\ \eta_m \end{array} \right )=0,
\label{tbmc}
\end{eqnarray*}
where
\begin{eqnarray*}
U_m&=&2\Lambda \cos(2\pi\alpha_1 m - 2\pi k_y),\\
V_m&=&2\Lambda \cos(2\pi\alpha_2 m - 2\pi k_y).
\end{eqnarray*}

For $\alpha_1=\alpha_2$ (mod 1), we recover the Abelian limit 
described by the Harper equation~\cite{Harper}
\begin{equation}
g_{m+1}+g_{m-1}+2\Lambda \cos(2\pi\alpha m - 2\pi k_y)g_m=E\,g_m,
\label{Harp}
\end{equation}
%where $\alpha=Bab/(hc/e)$ and $E(\alpha)$ is the energy in units of $-J$.
For irrational values of the flux $\alpha$, the system exhibits a metal-insulator transition at $\Lambda=1$.

The approach to the irrational values of $\alpha_i$ is studied by
considering a sequence of periodic systems obtained by rational
approximants $\alpha_i=p_i/q_i$. This
corresponds to truncating the continued fractional expansion of $\alpha_1$ and $\alpha_2$.
The resulting periodic system will have period $Q$, 
the least common multiple of $q_1$ and $q_2$. 
%For the golden mean, rational approximants are the ratio of two successive
%Fibonacci numbers $F_n$: $F_1=1$, $F_2=1$, $F_{n+1}=F_n+F_{n-1}$.

The $2Q$-dimensional system can be cast in the form
of two $Q$-dimensional eigenvalue problems:

\begin{eqnarray*}
\left( \begin{array}{cccccccc}U_1 & 1 & 0 & 0 &. &. &. & e^{-ik_A}\\
1 & V_2 & 1 & 0& 0 & .& .& 0\\
0 & 1 & U_3 & 1 & 0 & .& .& 0\\
0& 0& 1 & V_4 & 1 & 0 & .& 0\\
0& 0& 0& 1 & U_5 & 1 & .& 0 \\
.& .& .& .& .& .& .& .\\
e^{ik_A} & . & .& 0& 0& 0&1& V_Q \end{array} \right) \:
\left( \begin{array}{cc}\theta_1\\\eta_2\\ \theta_3\\ \eta_4\\.\\.\\.\\ \eta_Q\\ \end{array}\right) \: 
= E_A \left( \begin{array}{cc}\theta_1\\\eta_2\\ \theta_3\\ \eta_4\\.\\.\\.\\ \eta_Q\end{array}\right) \:
%\label{tbma]
\end{eqnarray*}
and
\begin{eqnarray*}
\left( \begin{array}{cccccccc}U_2 & 1 & 0 & 0 &. &. &. & e^{-ik_B}\\
1 & V_3 & 1 & 0& 0 & .& .& 0\\
0 & 1 & U_4 & 1 & 0 & .& .& 0\\ 
0& 0& 1 & V_5 & 1 & 0 & .& 0\\
0& 0& 0& 1 & U_6 & 1 & .& 0 \\
.& .& .& .& .& .& .& .\\
e^{ik_B} & . & .& 0& 0& 0&1& V_1 \end{array} \right) \:
\left( \begin{array}{cc}\theta_2\\\eta_3\\ \theta_4\\ \eta_5\\.\\.\\.\\ \eta_1\\ \end{array}\right) \:
= E_B \left( \begin{array}{cc}\theta_2\\\eta_3\\ \theta_4\\ \eta_5\\.\\.\\.\\ \eta_1\end{array}\right) \:
\label{tbmb}
\end{eqnarray*}
Here $E_A$ and $E_B$ denote the two sets of eigenvalues of the two uncoupled systems.
The allowed eigenenergies of the full system are the union of these two sets.

In the above two eigenvalue equations, we have used the Bloch condition,
\begin{eqnarray*}
\left( \begin{array}{cc}\theta_{2m-1+Q}\\ \eta_{2m+Q}\end{array}\right)\:
&=& e^{ik_AQ} \left(\begin{array}{cc} \theta_{2m-1}\\ \eta_{2m} \end{array} \right),\\
\left( \begin{array}{cc}\theta_{2m+Q}\\ \eta_{2m+1+Q}\end{array}\right)\:
&=& e^{ik_BQ} \left(\begin{array}{cc} \theta_{2m}\\ \eta_{2m+1} \end{array} \right ).
\end{eqnarray*}

\begin{figure}[htbp]
\includegraphics[width =1.0\linewidth,height=.7\linewidth]{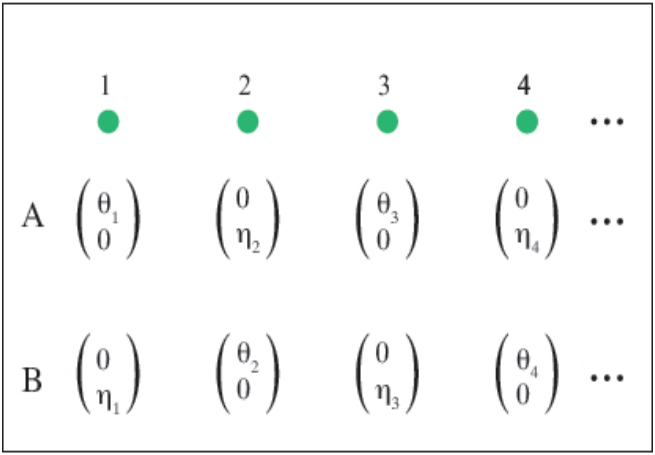}
\leavevmode \caption{(color online) The two possible antiferrimagnetic states designated as A and B.
}
\label{fig: AF}
\end{figure}

An important consequence of this decoupling of the $2Q$-dimensional
problem into two $Q$-dimensional problems is that the eigenstates of the system are of an
 ``antiferrimagnetic" type, as shown schematically 
in Fig.~\ref{fig: AF}. We will refer to them as of the A and B-type. 
The corresponding states denoted as $\chi_A$ and $\chi_B$
are in general non-degenerate.

It is instructive to compare the eigenvalue formulation to the transfer matrix approach 
discussed in earlier studies~\cite{NA}. The TBM equation can be written as a transfer matrix
equation,

\begin{eqnarray*}
\left( \begin{array}{cc} \theta_{m+1}\\ \eta_{m+1}\\ \theta_m\\ \eta_m \end{array}\right)\:
= T(m)  \left(\begin{array}{cc} \theta_m \\ \eta_m \\  \theta_{m-1} \\  \eta_{m-1}  \end{array} \right ),
\end{eqnarray*}
where
\begin{eqnarray*}
T(m) = \left(\begin{array}{cccc} 0&(E-V_m) & -1&0 \\ (E-U_m) & 0& 0& -1\\
1&0&0&0\\0& 1 &0 & 0 \end{array} \right ).
\end{eqnarray*}

The allowed energies are those for which the product of $Q$ successive matrices $B(m)$ has
an eigenvalue on the unit circle.
% which we denote as 
$e^{\pm i\phi_1}$ and $e^{\pm i\phi_2}$. 
Alternatively, the 4-dimensional 
transfer matrix equation can be reduced to two independent 2-dimensional transfer matrices as

\begin{eqnarray} 
\left( \begin{array}{cc} \theta_{m+2}\\ \eta_{m+1}\\ \end{array}\right)\:
&=& T_A(m)  \left(\begin{array}{cc} \theta_{m} \\ \eta_{m-1} \\  \end{array} \right ),\nonumber\\
\left( \begin{array}{cc} \theta_{m+1}\\ \eta_{m}\\ \end{array}\right)\:
&=& T_B(m)  \left(\begin{array}{cc} \theta_{m-1} \\ \eta_{m-1} \\  \end{array} \right ),
\label{matrix}
\end{eqnarray}
where the 2x2 matrices $T_A$ and $T_B$ are given by,
\begin{eqnarray*}
T_A(m)=\left( \begin{array}{cc} (E-V_{m+1})(E-U_{m-1})-1 & - (E-V_{m+1})\\  (E-U_{m}) & -1 \\ \end{array}\right)\:
\end{eqnarray*}
and
\begin{eqnarray*}
T_B(m)=\left( \begin{array}{cc} (E-V_{m})(E-U_{m-2})-1 & - (E-U_{m})\\  (E-V_{m-1}) & -1 \\ \end{array}\right).\:
\end{eqnarray*}

This decoupling of the $4$-dimensional transfer matrix problem into two $2$-dimensional
transfer matrices is equivalent to the decoupling discussed earlier for the eigenvalue problem,
which in turn implies the possibility of ``antiferrimagnetic" type states as shown in Fig.~\ref{fig: AF}.
An important consequence of this type of state is that (out of four), only two of the eigenvalues of the
$4$-dimensional transfer matrix have to be on the unit circle. In other words, in 
contrast to the statement
made in an earlier paper~\cite{NA}, the allowed
energies include states where two of the four eigenvalues of the transfer matrix are
not on the unit circle~\cite{prl}. 

The existence of ``antiferrimagnetic" states and the
relationship between the direct diagonalization method and the transfer matrix approach can be
illustrated by considering a simple non-Abelian system, namely the one 
with $\alpha_1=1/2$ and $\alpha_2=0$ which can be treated analytically.

%\begin{figure}[htbp]
%\includegraphics[width =1.1\linewidth,height=1.1\linewidth]{al2.eps}
%\leavevmode \caption{(color online) Energy spectrum viewed as a function of $k_y$ for $\alpha_1=\frac{1}{2}$
%,$\alpha_2=0$
%Note that the red regime actually corresponds to the Abelian problem with $\alpha_1=\alpha_2=1/2$,
%where the dispersion is given by $E=\pm \sqrt(\Lambda cos^2(k_y)+\Lambda cos^2(k))$}.
%\label{fig:E.2.0}
%\end{figure}

Diagonalization of two independent 2x2 matrices Eq.~(\ref{matrix}) gives
\begin{eqnarray*}
E_A(k_A,k_y)&=&2\left[\Lambda \cos 2\pi k_y \pm \cos k_A\right],\\
E_B(k_B,k_y)&=&\pm 2 \sqrt{\Lambda^2 \cos^2 2\pi k_y+\cos^2 k_B},
\end{eqnarray*}
and the corresponding eigenvectors
\begin{eqnarray*}
\chi_A=\left( \begin{array}{cc} \pm e^{-ik_A}\\0\\ 1\\ 0 \end{array}\right), \:
\chi_B=\left(\begin{array}{cc} 0\\ e^{-ik_B}\frac{E/2 \pm \Lambda \cos 2\pi k_y}{\cos k_B}\\0\\1 \end{array} \right ).
\end{eqnarray*}

%In this simple system, the A and B-type solutions respectively correspond to the two Abelian
%systems with $\alpha_1=\alpha_2=0,1/2$.
In general, the A and the B-states are non-degenerate.
As is explicit in this example, the magnitude of the two components of the vectors are
in general unequal, and hence the solutions correspond to antiferrimagnetic states.
However, at $k_y=\pi/2$, the $E_A=E_B$ and the two degenerate states are
antiferrimagnetic.

The spectrum can also be obtained by iterating the transfer matrix problem
where the energies can be written in terms of the eigenvalues of the transfer matrix, $e^{\pm i \phi_A}, e^{\pm i\phi_B}$.
Comparison of the spectrum obtained by these two methods shows that $\phi_A=k_A/2$
and $\phi_B=k_B/2$.

The results of this paper were obtained using the direct diagonalzation method.
We shall henceforth set $\alpha_1 = \gamma$, the golden mean, and explore various 
complexities of the problem for different values of $\alpha_2$.

\section{Non-Abelian Spectrum for Irrational $\alpha_1$ }

The energy spectrum of the system 
is the union over $k_y$ of the individual energy spectra of the A and the B types of 
the tight binding equations. In the Abelian case,
$\alpha_1=\alpha_2$, and when $\alpha_1$ is rational, equal to $p/q$, the spectrum consists of $q$
bands which are usually separated by gaps. As $k_y$ varies, the bands shift and their width may change,
but they do not overlap, except at the band edges. For irrational
$\alpha$, the spectrum is independent of $k_y$.

\begin{figure}[htbp]
\includegraphics[width =1.1\linewidth,height=1.1\linewidth]{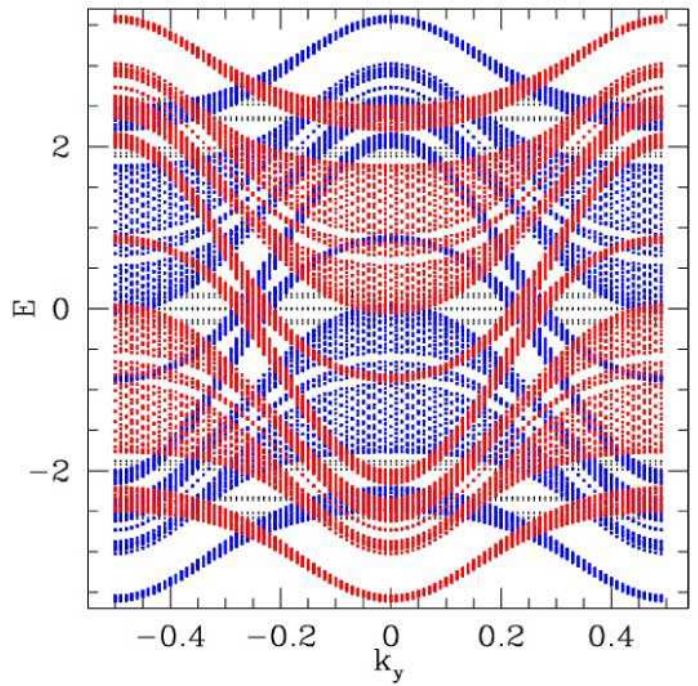}
\leavevmode \caption{(color online) Energy spectrum viewed as a function of $k_y$ for
$\alpha_1=89/144$ and $\alpha_2=\frac{1}{2}$ with $\Lambda=1$ for a range of $k_x$ values.
The red and blue correspond to $E_A$ and $E_B$ respectively.
The grey dots show the corresponding Abelian case with $\alpha_1=\alpha_2=\gamma$.}
\label{Eg1.2}
\end{figure}

\begin{figure}[htbp]
\includegraphics[width =1.1\linewidth,height=1.1\linewidth]{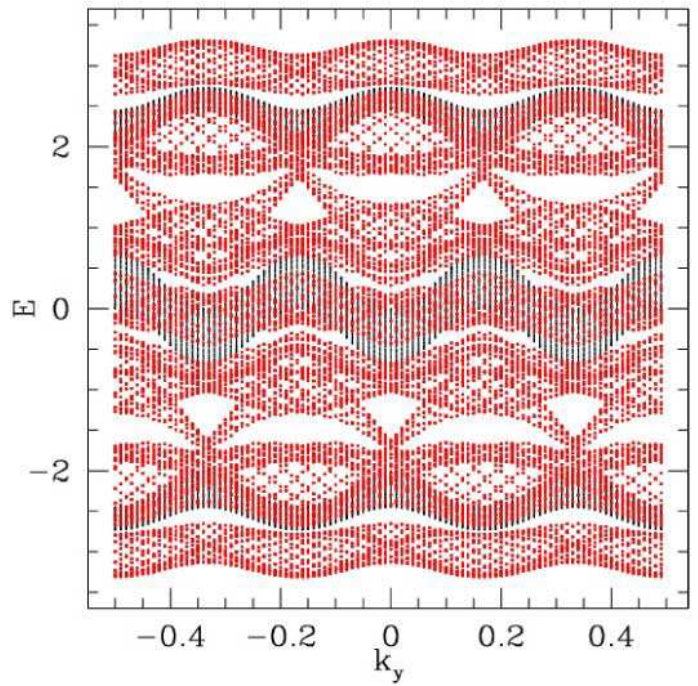}
\leavevmode \caption{(color online) Energy spectrum viewed as a function of $k_y$ for
$\alpha_1=89/144$ and $\alpha_2=\frac{1}{3}$ with $\Lambda=0.5$.
The grey dots show a corresponding Abelian case with $\alpha_1=\alpha_2=\frac{1}{3}$.}
\label{E1.3.3}
\end{figure}

A striking aspect of the non-Abelian problem is the overlapping of the bands,
as illustrated in figures~\ref{Eg1.2} and ~\ref{E1.3.3}.
These figures depict two different classes of typical non-Abelian spectra
with rational $\alpha_2$:
In Fig. ~\ref{Eg1.2}, type A and B solutions result in 
a non-degenerate spectrum; and
Fig. ~\ref{E1.3.3} shows the case where A and B solutions are degenerate.
Below we discuss various spectral characteristics of the system.

For rational $\alpha_2=p_2/q_2$, the spectrum is a periodic function of $k_y$.
This is due to the fact that for irrational $\alpha_1$, the set $\{U_m\}$ is 
ergodic in $m$, while the set $\{V_m\}$ is periodic in $m$ for rational $\alpha_2$.
We list below some of the characteristic properties of the spectrum:
\begin{description}
\item{(1)} $E_{A,B}(2\pi k_y)=-E_{A,B}(2\pi k_y+2\pi\alpha_2)$,
\item{(2)} $E_A(2\pi k_y)=-E_B(2\pi k_y+2\pi\alpha_2)$,
\item{(3)} $E_{A,B}(2\pi k_y)=E_{A,B}(2\pi k_y+4\pi\alpha_2)$,
\item{(4)} $E(2\pi k_y)=E(2\pi k_y+4\pi\alpha_2)$.
\end{description}

For certain values of $\alpha_2$,  A and B-type of states
are degenerate.
This happens when
the two sets $V_{2m}(k_y)$ and $V_{2m+1}(-k_y)$ coincide.
This degeneracy occurs when (a) $\alpha_2$ is an irrational number and (b)
$\alpha_2=p/q$ with $q$-odd as $V_{m+qn}(k_y)=V_{q-m+qn}(-k_y)$ (where $n$ is an integer and $m < q$). 
%That is, odd-$q$ connects $V_m$ at even and odd-sites. The $q=3$ is a special case
%where $V_m(k_y)=V_{m+1}(-k_y)$, while for other values of $q$, the two sets $V_{2m}$ and $V_{2m+1}$
%become identical even though the $V_m$ at neighboring sites is different.
For even $q$, the A and B states are in general non-degenerate.
This distinction between the odd and the even cases leads to significant differences between these two cases.

\section { The Localization Transition }

The metal-insulator transition\cite{Sok} in 2DEGs in the presence of a magnetic (Abelian) field
is a paradigm for the Anderson localization transition.
We now discuss the corresponding localization transition that exists in the non-Abelian systems.
In contrast to the Abelian case, where all states localize at the same value of the
tunneling anisotropy, localization
in the non-Abelian case varies throughout the spectrum.

In this section, we will discuss the localization properties of the $E=0$ state
a study is relevant for fermionic atoms near half-filling.
As shown below, for $\alpha_2=\frac{1}{2}$ as well as for $\alpha_2=\frac{1}{4}$,
the onset to a localization transition can be inferred from the well-known localization characteristics of the 
Harper equation.

\bigskip
{\bf A1: Localization Boundary for $\alpha_2=1/2$}\\

\begin{figure}[htbp]
\includegraphics[width =1.1\linewidth,height=1.1\linewidth]{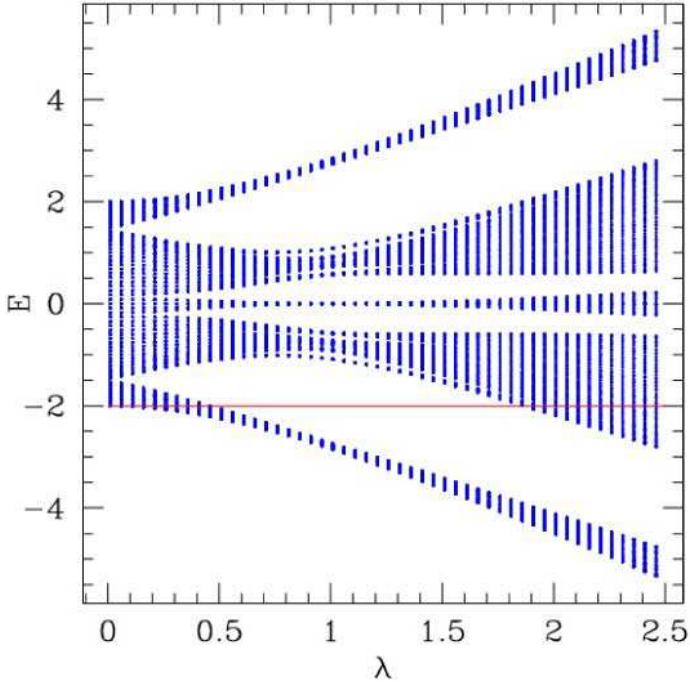}
\leavevmode \caption{(color online) The eigenvalues of the effective TBM
(Eq.~(\ref{Hlike})) describing $E=0$ states as $\lambda$ varies. The line (red online) shows the
values of $\lambda$ where $E=0$ is an eigenvalue of the TBM (Eq. \ref{Hlike}) }
\label{fig:harp}
\end{figure}

\begin{figure}[htbp]
\includegraphics[width =1.1\linewidth,height=1.1\linewidth]{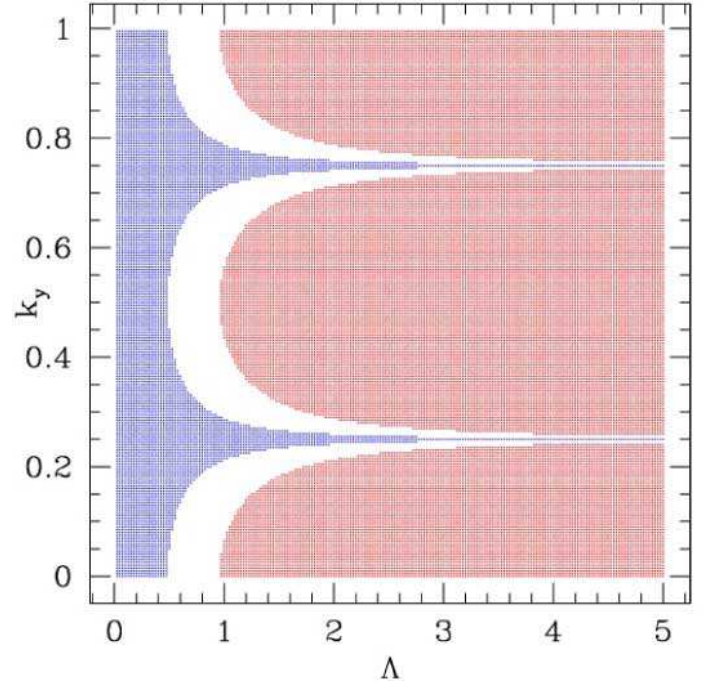}
\leavevmode \caption{(color online) With $\alpha_2=1/2$ shaded regime (blue) shows
the extended phase while shaded red shows the localized phase of $E=0$ state in $\Lambda-k_y$ plane.}
\label{fig:E0loc.2}
\end{figure}

For $\alpha_2=\frac{1}{2}$, the coupled TBM equations (Eq. ~\ref{Hlike}) for $E=0$ reduce to
\begin{eqnarray*}
\theta_{m+2}+\theta_{m-2}+2(-1)^m \lambda \cos(2\pi\alpha_1 m -2\pi k_y)\theta_m&=& \epsilon \theta_m\\
\eta_{m+1}+\eta_{m-1}+2\Lambda \cos(2\pi\alpha_1 m -2\pi k_y)\theta_m&=& 0 \nonumber
\label{Hlike}
\end{eqnarray*}
where $\epsilon=-2$ and 
$\lambda=2\Lambda^2 \cos(2\pi k_y)$.

For $E=0$, the uncoupled $\theta$-equation maps to
an $E=-2$ Harper-like equation 
(Eq.~(\ref{Hlike})), where the on-site quasiperiodic potential is a sinusoidal 
function of $k_y$. The eigenstates of this system localize at $\lambda=1$, providing an 
explicit threshold for localization of the $E=0$ state of the non-Abelian system 
provided $\epsilon=-2$ is the eigenvalue of Eq.~(\ref{Hlike}).

As shown in Fig.~\ref{fig:harp}, $\epsilon=-2$ is an eigenvalue of the system provided 
$\lambda=\lambda_1 \lesssim 0.48$ or $\lambda=\lambda_2 \ge 1.83$.
These critical values determine the boundary curves for the localization of the $E=0$ 
state: in the Harper equation, all states are extended 
for values of $\lambda \le 1$. These two localization boundaries are exhibited in 
Fig. ~\ref{locb.2}
\\
\\
{\bf A2: Localization Boundary for $\alpha_2=1/4$ }\\

\begin{figure}[htbp]
\includegraphics[width =1.1\linewidth,height=1.1\linewidth]{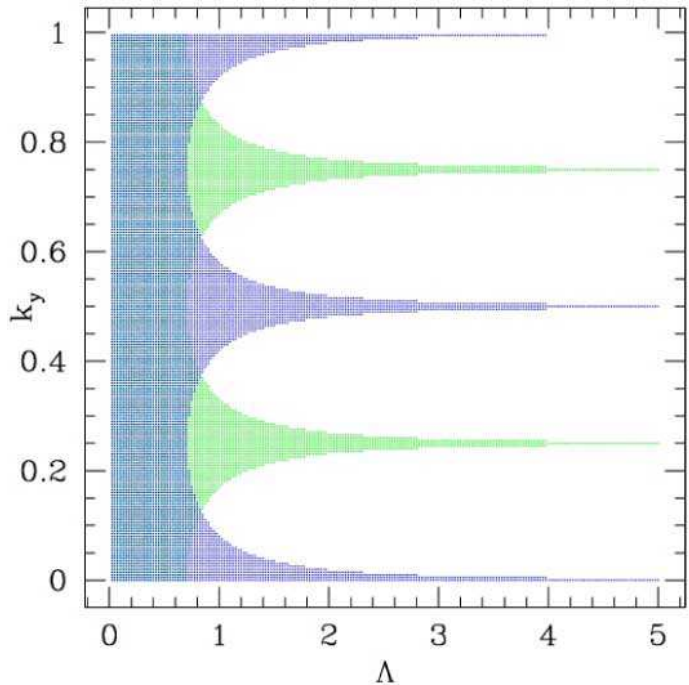}
\leavevmode \caption{(color online) For $\alpha_2=1/4$, shaded regimes show extended phase for
two degenerate $E=0$ states
belonging to the A (green) and B (blue) sectors in $\Lambda-k_y$ plane. }
\label{fig:E0loc.4}
\end{figure}

For $\alpha_2=\frac{1}{4}$, the uncoupled $\theta$-equations
for the A and the B-sectors of the TBM for $E=0$ reduce to
\begin{eqnarray*}
\bar{\theta}^A_{m+2}+\bar{\theta}^A_{m-2}+2 i \lambda_A \cos( 2\pi\alpha_1 m-2\pi k_y)\bar{\theta}^A_m&=&0,\\
\bar{\theta}^B_{m+2}+\bar{\theta}^B_{m-2}+2 i \lambda_B \sin( 2\pi\alpha_1 m-2\pi k_y)\bar{\theta}^B_m&=&0,
\end{eqnarray*}
where 
\begin{eqnarray*}
\lambda_A&=&2\Lambda^2 \cos(2\pi k_y),\\
\lambda_B&=&2\Lambda^2 \sin(2\pi k_y).
\end{eqnarray*}
The above two equations correspond to A and B-type states with $E=0$, respectively.
Here $\bar{\theta}_m^{A,B}=i^m \theta_m^{A,B}$.  In a manner analogous to the corresponding Hermitian problem, 
the system exhibits self-duality at $\lambda_{A,B}=1$ and this self-dual point describes the onset of 
localization~\cite{Amin}.  For $k_y=\frac{1}{8}$~(mod $\frac{1}{4}$), both types of solutions
localize simultaneously. However, at other values of the transverse momentum, only one
of the states is localized. This is an example of two degenerate states with different
transport properties:  depending upon $(\Lambda,k_y)$, type A states may be extended (localized) while type B 
states will be localized (extended).
This localization boundary in $\Lambda-k_y$ space is shown
for types A and B in Fig.~\ref{fig:E0loc.4}.

The existence of conducting states for all values of $\Lambda$ is one of the most intriguing
characteristics of the non-Abelian system. Below we show that these states defying localization
describe relativistic particles.

\bigskip
%{\bf B: Dirac Fermions}\\
{\bf B: Relativistic Dispersion and Defiance of Localization}\\

\begin{figure}[htbp]
\includegraphics[width =1.1\linewidth,height=1.1\linewidth]{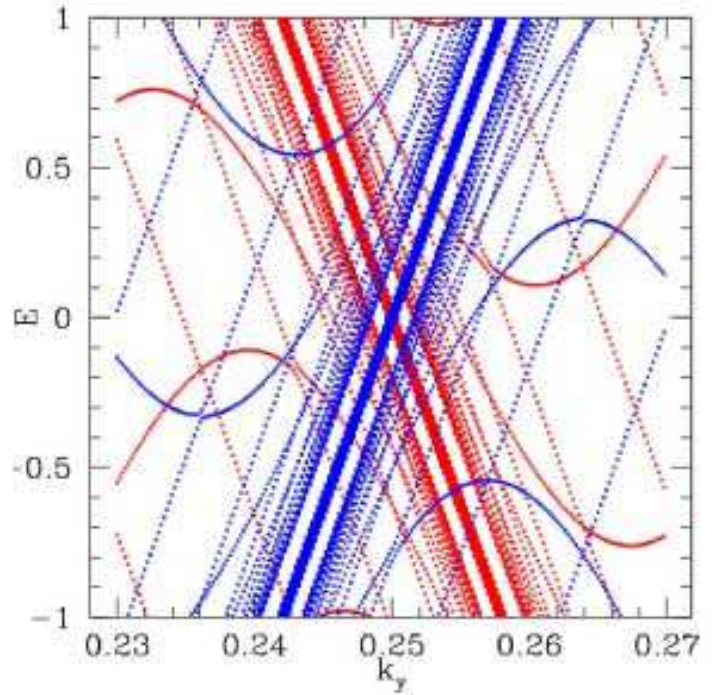}
\leavevmode \caption{(color online) Blowup of the $E=0$ and $k_y=1/4$ neighborhood for
$\alpha_1=89/144$ and $\alpha_2=\frac{1}{2}$ and $\Lambda=10$.
The red and blue respectively correspond to $E_a$ and $E_b$.}
\label{fig:E10.2}
\end{figure}

Figure~\ref{fig:E10.2} shows the energy-momentum relation for $\alpha_2=1/2$
near $E=0, k_y=1/4$.
Although the level  structure is complicated,
near $k_y=1/4$, the energy bands
exhibit the linear dispersion characteristic of the one-dimensional relativistic particles.
Thus, the non-Abelian system with A and B type states,
provides an interesting manifestation of the positive and the negative energy
states of a one-dimensional relativistic particle.

An important characteristic of the states that reside at the crossings is that they defy localization.
It should be noted that a crossing at $E=0$ exists irrespective of the value of $\Lambda$. 
In other words, we have a relativistic theory
for all values of $\Lambda$ as shown in the Figures~\ref{Eg1.2} and~\ref{fig:E10.2}.
Such states remain extended irrespective of the quasiperiodic disorder in the system
as the linear dispersion exists for the full range of $\Lambda$ values.

For $\alpha_2=\frac{1}{4}$, we obtain an effective relativistic theory
for zero-energy states near $k_y=0$ for type-A and near $k_y=\pi/2$ for type-B states.
These states remain conducting for all values of $\Lambda$.

We would like to note that
in the Abelian system, a linear energy-momentum relation resulting in a Dirac cone
occurs for rational values of $\alpha$ in the neighborhood of some special values of $k_x,k_y$ near $E=0$.
However, for irrational $\alpha$, the spectrum is independent of $k_y$ and the Dirac cone disappears.
Therefore, in the Abelian case, effective relativistic theory bears no relationship to the transport properties
as the states are always extended for rational $\alpha$.

\bigskip
{\bf C: Localization Transition and Loss of Relativistic Dispersion}\\

Our detailed investigation for various values of $\alpha_2$ shows that the presence of conducting states for all values of $\Lambda$ is not a generic property of the system. In particular, for cases where the type-A and type-B states are always
degenerate, all states are found to localize. Interestingly, the transition to localization
is accompanied by a loss of the relativistic character of the energy momentum relation.

\begin{figure}[htbp]\includegraphics[width =1.1\linewidth,height=1.1\linewidth]{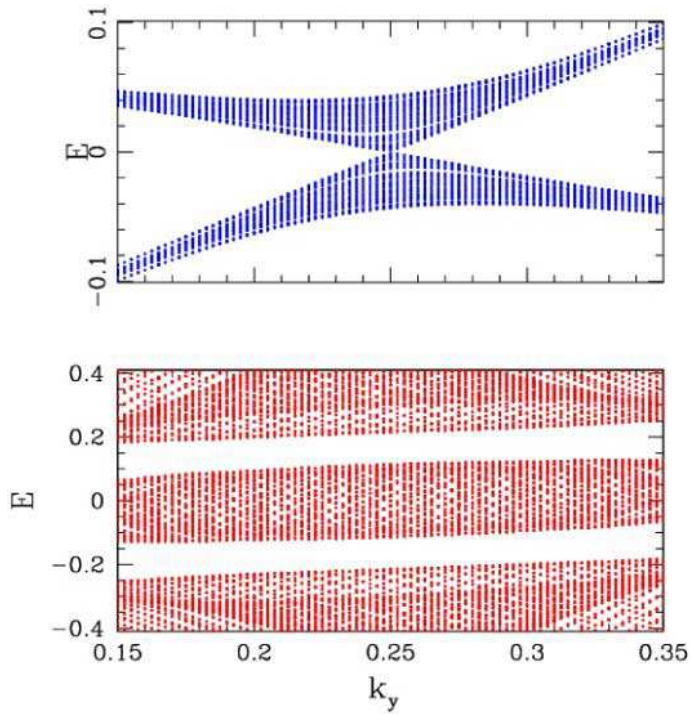}
\leavevmode \caption{(color online) Top and bottom panel respectively show the
spectrum for $\alpha_2=\gamma^3$ with $\Lambda=0.75, 1.25$ which respectively correspond to
extended and localized $E=0$ states.}
\label{E0.g3}\end{figure}

\begin{figure}[htbp]
\includegraphics[width =1.0\linewidth,height=1.1\linewidth]{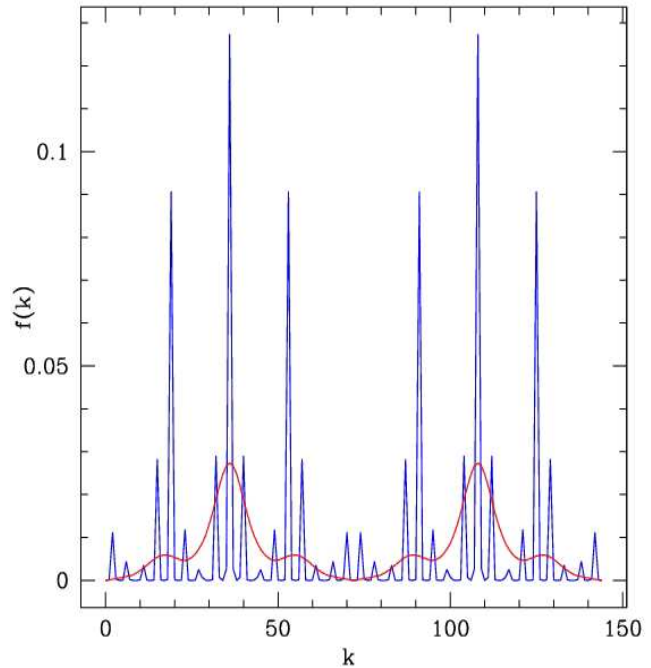}
\leavevmode \caption{(color online) Fourier transform of the wave function for $E=0$ state with
 $\alpha_2=\gamma^3$ with $\Lambda=0.75$ (sharp fringes)
and $\Lambda=1.25$ (smeared out fringes). The label $k$ on the x-axis corresponds to the momentum
$2\pi k/L$, where $L$ is the size of the lattice.}
\label{FT.g3}
\end{figure}

For example, for $\alpha_2=p/q$ where $q$ is odd, as well as for irrational $\alpha_2$,
the crossings characterizing certain $E=0$ states disappear beyond a certain critical value of $\Lambda$.
Interestingly, this threshold for the disappearance of the crossing is always
found to coincide with the onset to localization of that state.
Figures~\ref{E0.g3} and~\ref{FT.g3} illustrate this for irrational $\alpha_2$ as the disappearance of band crossings is accompanied by
the broadening and flattening of the Bragg peaks.

\newpage

\section { Localization Transition of Bose-Einstein Condensates }

The natural locus for BEC in ultracold atoms in optical lattices is the band edge.
We now explore the spectral and transport properties of the states at band edges, namely
the minimum energy states as $k_y$ varies.

In contrast to the preceding analytical treatment of the band centers,
we have investigated localization properties of the states at the band edge
with numerical methods.

\begin{figure}[htbp]
\includegraphics[width=1.0\linewidth,height=1.1\linewidth]{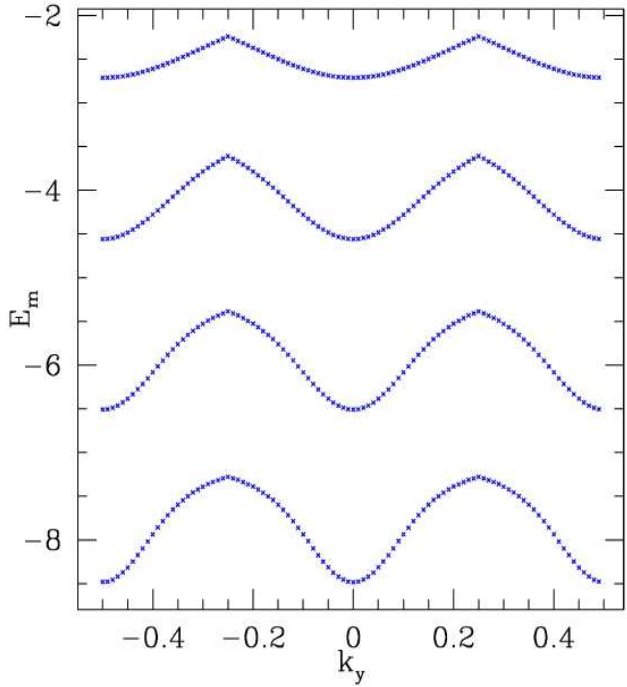}
\leavevmode \caption{(color online) Minimum energy as a function of $k_y$
for $\alpha_2=\frac{1}{2}$ with $\Lambda=0.5,1.5,2.5,3.5$ (top-bottom) illustrating
the change from linear to quadratic dispersion near $k_y=\pm 1/4$.}
\label{Emin.2}
\end{figure}

As seen from Figs~\ref{Eg1.2} and \ref{Emin.2}, the energy spectrum for $\alpha_2=\frac{1}{2}$ shows
the existence of a linear dispersion relation near the band crossings. As the lattice anisotropy $\Lambda$ varies, we see a transition from relativistic to non-relativistic behavior near $\Lambda \approx 2.5$; this transition is accompanied by the loss of the wave function's spinor character, causing an effective spin polarization.

\begin{figure}[htbp]
\includegraphics[width=1.0\linewidth,height=1.3\linewidth]{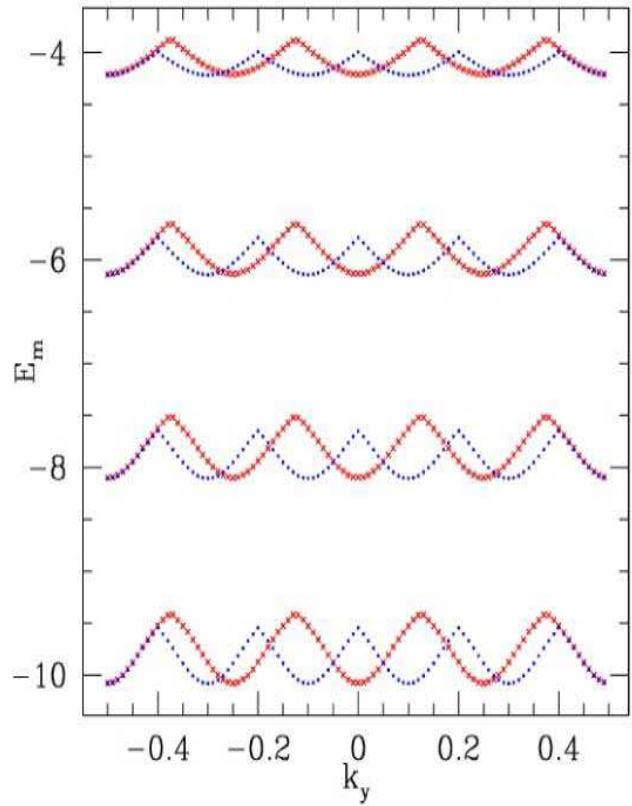}
\leavevmode \caption{(color online) Minimum energy as a function of $k_y$
for $\alpha_2=1/4$ (cross), $\alpha_2=1/5$ (dots) with $\Lambda=1.5,2.5,3.5, 4.5$ (top-bottom) illustrating
the change from the linear to quadratic dispersion near $k_y=\pm 1/4$.}
\label{Emin.4.5}
\end{figure}

\begin{figure}[htbp]
\includegraphics[width=1.1\linewidth,height=1.5\linewidth]{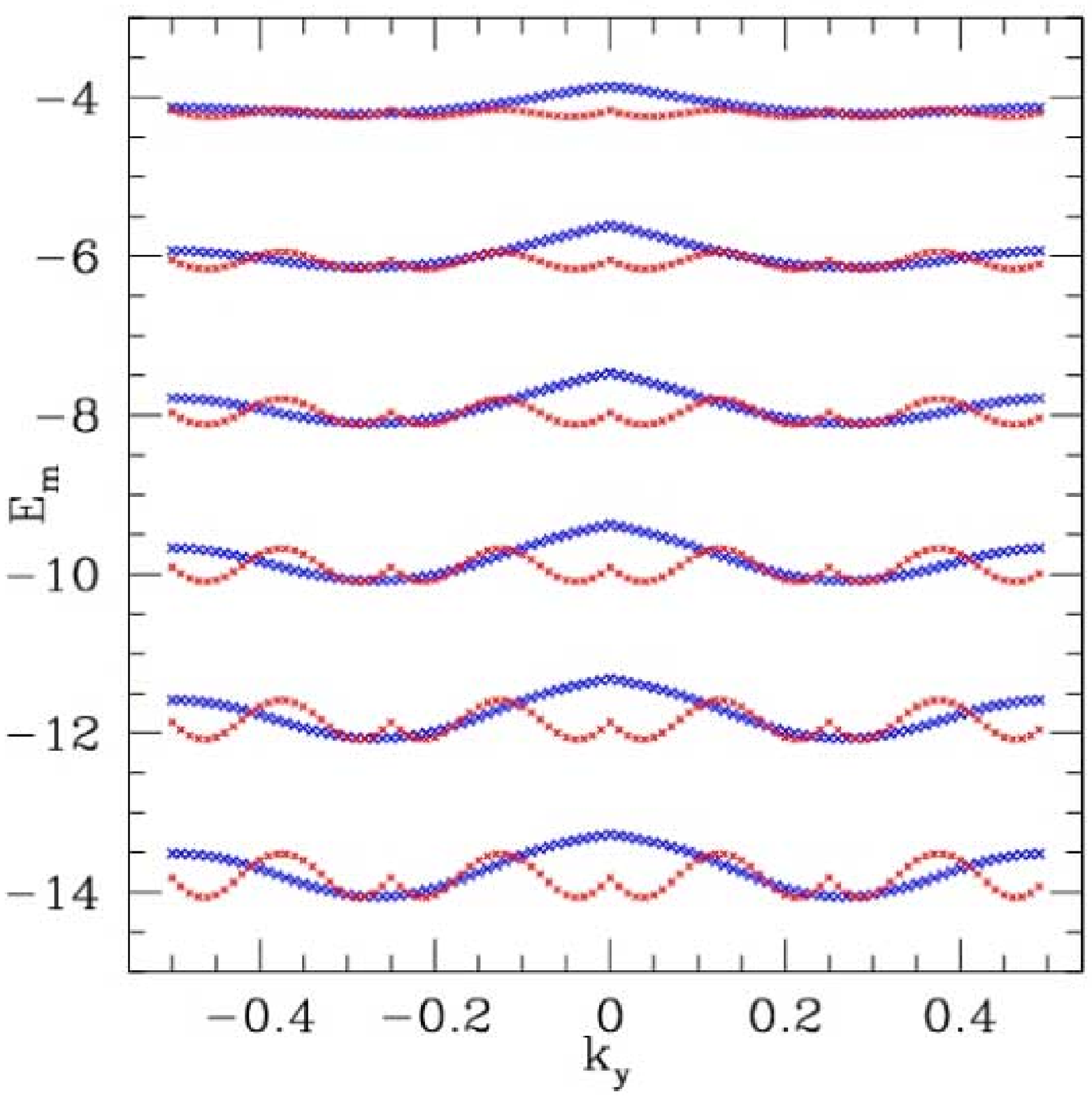}
\leavevmode \caption{(color online) Variation of minimum energy with $k_y$
for $\alpha_2=\gamma^3$ (large crosses) and $\alpha_2=\gamma^4$ (smaller crosses) with $\Lambda=1.5,2.5,3.5,4.5,5.5,6.5$ (top-bottom).}
\label{Emin.g3.g4}
\end{figure}

The robustness of the linear dispersion in non-Abelian systems is shown
for various values of $\alpha_2$ in Figs. ~\ref{Emin.4.5} and ~\ref{Emin.g3.g4}.
It appears that it is only in the even-$q$ cases that the nature of the dispersion changes as $\Lambda$ varies.
Similarly, for $\alpha_2=\gamma^4$ (an odd harmonic of $\gamma$, as $\gamma^4=2-3\gamma$), linear dispersion at $k_y=0$ and at $k_y=1/4$ occurs for all values of $\Lambda$, while for $\alpha_2=\gamma^3$ (an even harmonic of $\gamma$, as $\gamma^3=2\gamma-1$), a relativistic energy-momentum relation is seen
for small and large values of $\Lambda$ as illustrated in Fig.~\ref{Emin.g3.g4}.
In other words, a ``transition" from relativistic to non-relativistic behavior
can be induced by varying $\Lambda$ for some values of $\alpha_2$.

Another point to be noted is that the ground state of the system may have nonzero momentum:  for even $q$, the global energy minimum occurs at $k_y=0$, while for odd $q$, it occurs at $k_y=\alpha_2/2$.

\begin{figure}[htbp]
\includegraphics[width =1.0\linewidth,height=1.5\linewidth]{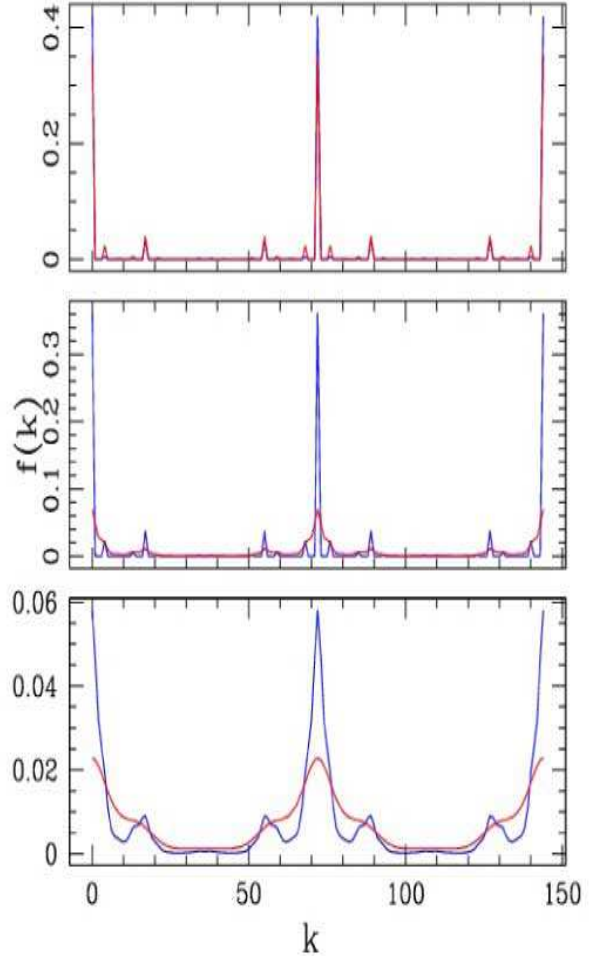}
\leavevmode \caption{(color online) Fourier transform of the wave function for $\Lambda=.4,.5,.7$ (top-bottom).
Each caption shows $k_y=.25$(red) and $k_y=0$(blue) for the minimum energy state for $\alpha_2=1/2$.
The x-axis index $k$ corresponds to the Bloch vector $2\pi k/L$. There are peaks corresponding to the irrational values of $\alpha_1$,
occuring at $k$ values equal to half of the Fibonacci numbers (as anti-ferromagnetic nature effectively doubles the size of the unit cell).
Additional satellite peaks characterize the non-Abelian ferature of the system.}
\label{FTT}
\end{figure}

Figures~\ref{FTT} illustrate the localization transition of the minimum energy states.
Extended states in these figures are characterized by sharp Bragg peaks in the momentum distribution, 
and the localization transition is signaled by the broadening of these peaks.  As we increase the parameter $\Lambda$, the $k_y=\pm 1/4$ states localize before the $k_y=0$ state. Our detailed investigation shows that $k_y=0$ is the last state to localize as $\Lambda$ is varied for all values of $\alpha_2$. This is contrary to familiar experience, in which localization begins at the band edge. The localization for the minimum energy state is insensitive to the energy-momentum relation, in contrast to the $E=0$ states.

For irrational $\alpha_2$, we expect the localization threshold to be lowered.  Our numerical results show that the minimum energy states begin to localize at a relatively small value of $\Lambda \approx 0.15$. As discussed earlier, $E=0$ states resist localization due to their linear dispersion but eventually localize. Our numerical studies show that
localization is complete at $\Lambda=1$, as in the Abelian case.

\newpage

\section{Experimental Observation of Metal-Insulator Transition }

\begin{figure}[htbp]
\includegraphics[width =1.0\linewidth,height=1.1\linewidth]{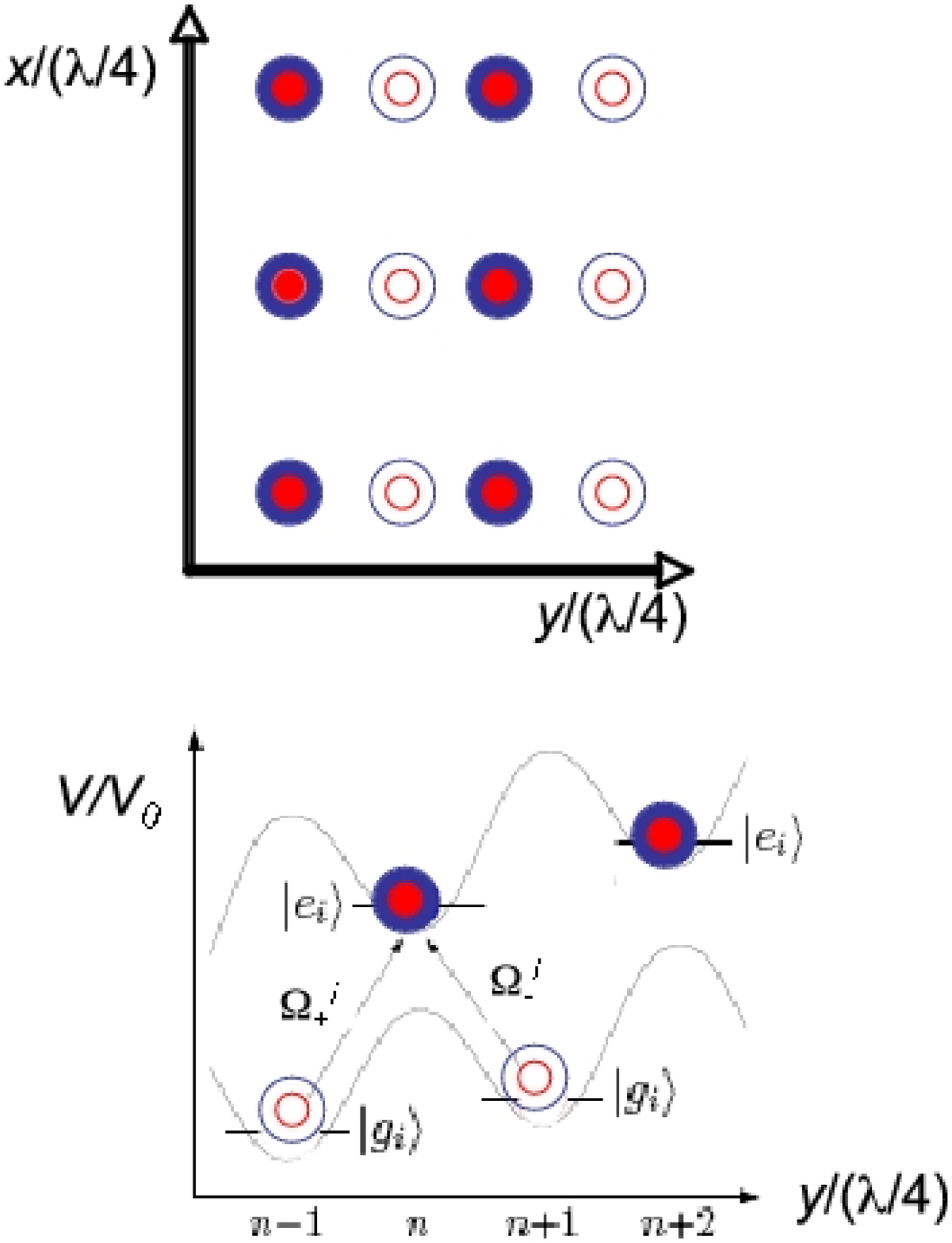}
\leavevmode \caption{(color online) Schematic diagram illustrating the non-Abelian $U(2)$ set up. The ground states are hollow and the excited states are filled. Red and blue represent the two ``colors" of the $U(2)$ group.}
\label{naExp}
\end{figure}

An experimental setup for generating artificial Abelian and non-Abelian fields consists of~\cite{NJP, NA} a
two-dimensional optical lattice populated with cold atoms that occupy two hyperfine states.
The lattice laser polarization is adjusted to confine these states to alternating columns.
The non-Abelian scheme requires atoms with two pairs of hyperfine levels:  $|g_1\rangle$, $|e_1\rangle$, $|g_2\rangle$, $|e_2\rangle$ as shown in Fig.~\ref{naExp}.

The typical kinetic energy tunneling along the $y$-direction is suppressed by accelerating the system or
applying an inhomogeneous electric field in that direction such that the lattice is tilted. 
Tunneling is instead accomplished with two sets of laser-driven Raman transitions with space-dependent
Rabi couplings $\Omega_j e^{iq_j y}$ where $j = 1, 2$.
%induced by additional running wave lasers detuned by $\pm\Delta$ to cancel the tilting effect of the acceleration.
The wave numbers $q_j$ generate an effective magnetic flux where
$q_j=(2\pi \alpha_j) / a$, where $\lambda=2a$ is the wavelength of the laser light.
In an optical lattice with a finite number of sites,
the two components of the "magnetic flux" ( $\alpha_1,\alpha_2$ ) can be adjusted, in a controlled manner, 
to a sequence of rational approximants to the golden mean by tuning the $q_j$.
We direct readers to~Refs.\cite{NJP, NA} regarding various details for generating these artificial gauge fields.
\begin{figure}[htbp]                                                                                      
\includegraphics[width =0.9\linewidth,height=0.7\linewidth]{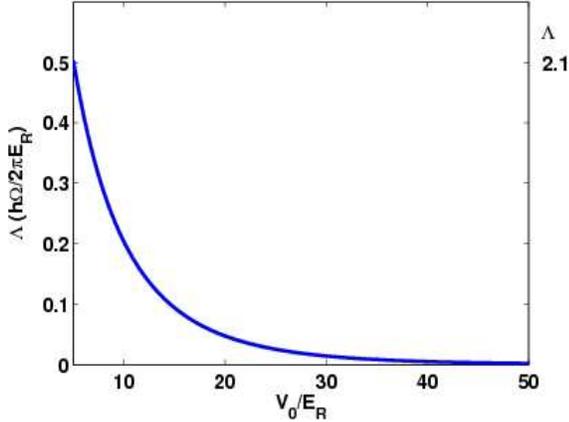}                                     
\leavevmode \caption{(color online) $\Lambda$                                                             
as the depth of the 2D optical lattice is tuned with $\alpha=\frac{1}{2}$.                                   
The factor $E_R/\hbar\Omega \approx 4.2$ with the following laser                                            
parameters: $I_g=1$~mW/cm$^2$, $x=11$, $\delta_r=100$~GHz, $E_R/h=3.2$~kHz.}
\label{trans}                                                                                             
\end{figure}         

We now describe the experimental feasibility of tuning $\Lambda$ to induce metal-insulator transitions by adjusting the lattice beam intensity $V_0$. For simplicity, we will initially discuss the Abelian case. Let us first consider the laser assisted coupling $J_y$ as a function of $V_0/E_R$, where $E_R=2\pi^2\hbar^2/M\lambda^2$ is the photon recoil energy.
The tunneling is defined as the matrix element of the
Rabi coupling ($\Omega$) between Wannier functions $w$, evaluated at the two adjacent lattice sites:
\begin{equation}
J_y=\int w(\vec{x}-\vec{x}_i) \frac{\hbar}{2}\Omega\exp(iqx)w(\vec{x}-\vec{x}_{i-1})d^3\vec{x},
\end{equation}
where $q = (2 \pi \alpha)/a$.
% where for the laser beam of wavelength $\lambda$,  we have $q=4 \pi \alpha/\lambda$.
The Wannier functions for
$V(x)=V_0\sin^2(2\pi x/\lambda)$] have been computed ~\cite{NJP}; $J_y$ decreases monotonically with $V_0/E_R$. 
This basic behavior can be demonstrated analytically by assuming a deep lattice approximated by a harmonic oscillator potential and taking the Wannier functions to have the corresponding Gaussian form. The Gaussian approximation yields
\begin{equation}
J_y = \frac{\hbar\Omega}{2} \exp[-\frac{\pi^2}{16}\sqrt{V_0/E_R}]\exp[-\frac{\alpha^2}{\sqrt{V_0/E_R}}].
\end{equation}
The kinetic energy coupling in the $x$-direction $J_x$
also decreases monotonically with $V_0/E_R$ for sufficiently large values~\cite{RPCW} as described by
\begin{equation}
J_x \approx 1.397E_R\left(\frac{V_0}{E_R}\right)^{1.051}\exp[-2.121\sqrt{V_0/E_R}].
\end{equation}

The ratio of $\Lambda=J_x/J_y=(E_R/\hbar\Omega)\,f(V_0/E_R,\alpha)$ is shown in Fig.~\ref{trans} for a characteristic range of $V_0/E_R$ with the scale set by the factor
$\hbar\Omega/E_R$. In order to generate a useful range of $\Lambda$ values ({\it e.g.},$0 < \Lambda \lesssim 2$), the parameter $\hbar\Omega/E_R$ must be set to order unity.

We now argue that it is possible to achieve this with reasonable experimental settings.  We consider the case of $^{87}$Rb, where the $|g\rangle$ and $|e\rangle$ states are taken to be the hyperfine levels of the $5^2S_{1/2}$ level~\cite{data} and the Raman level is $5^2P_{3/2}^\circ$.
The parameter $\hbar\Omega/E_R$ can be fixed near unity if the Raman laser beams have intensities on the order of 1 mW/cm$^2$ with $x\approx11$ and are detuned from the $^{87}$Rb~$D_2$ line by about $\delta_r\approx100$~GHz. 
We require that the Rabi coupling of the Raman transition $\Omega$, the detuning $\Delta$, 
and the lattice trapping frequency $\nu_x=\sqrt{4 E_R V_0}/ \hbar$ have well separated magnitudes such that $\Omega\ll\Delta\ll\nu_x$, to 
ensure that only the lowest band of the lattice is occupied and no other excitations occur.
Typical values of $\nu_x$ are on the order of tens of kilohertz.
The Raman transition is stimulated by two lasers with Rabi couplings $\Omega_g^{(1)}$ and $\Omega_e^{(1)}$ and intensities $I_g$ and $I_e$ with a
large detuning $\delta_r$ such that the effective Rabi coupling magnitude is $\Omega=\Omega_g^{(1)}\Omega_e^{(1)}/2\delta_r$. The Rabi coupling $\Omega$ can be written as a product of atomic factors and laser tuning parameters,
\begin{equation*}
\Omega=\left(\frac{\Gamma^2}{4 I_{sat}}\right)\left(\frac{\sqrt{\xi}I_g}{\delta_r}\right),
\end{equation*}
where $\Gamma$ is the natural decay rate of the $5^2P_{3/2}\circ$ state and $I_{sat}$ is the saturation intensity of the 
$D_2$ line (See Ref.~\cite{data}). The ratio $\xi=I_e/I_g$ must be less than 0.17 or greater 
than 5.8 to satisfy the $\Omega\ll\Delta$ condition.
%The parameter $\hbar\Omega/E_R$ can be fixed near unity if the Raman laser beams have intensities on the order of 1 mW/cm$^2$
%with $\xi\approx11$ and are detuned from the $^{87}$Rb~$D_2$ line by about $\delta_r\approx100$~GHz.
The separation of scales between the Rabi coupling $\Omega$ of the Raman transition and lattice trapping frequency $\nu_x$
necessary to generate the magnetic field in the above scheme (i.e., $\Omega\ll \nu_x$) is sufficient to generate a reasonable range of $\Lambda$ values.

In the non-Abelian case, there are generally two possible values of $\Lambda$ corresponding to $\Omega_1$ and $\Omega_2$, one for each ``color". By adjusting $\Omega_2/\Omega_1=f(V_0/E_R,\alpha_1)/f(V_0/E_R,\alpha_2)$, we obtain a single $\Lambda$ in correspondence with the theoretical studies described here.

\section{Summary}

This paper discusses spectral and transport properties of the cold atom analog of a 2DEG in a lattice, subject to a non-Abelian gauge field with $U(2)$ symmetry.

In the continuum limit of the lattice, the system maps onto two oppositely-charged coupled harmonic oscillators, with a coupling constant proportional to the strength of the non-Abelian field.  The Landau energy levels of the Abelian problem evolve into entangled states of this particle-hole pair.

These features also characterize the energy spectrum of the the corresponding lattice problem.  In fact, the transition from Landau levels to Landau bands is the analog of the generalization from the butterfly to the moth spectrum as the Abelian system becomes non-Abelian.  The non-Abelian coupling breaks the degeneracy of the Landau levels; the spectrum depends explicitly on the transverse momentum.

The non-Abelian system exhibits antiferrimagnetic-type ground states, whose components, A and B, need not be degenerate, and in fact may have very different transport properties. A particularly interesting example of this is the zero-energy state for $\alpha_2=1/4$, where the degenerate A and B components have different localization properties.  Additionally, an intriguing relationship between the A and the B components occurs for $\alpha_2=\frac{1}{2}$, as these two components correspond to the positive and the negative energy states of the system.  Such novelties may open new avenues for exploring frontiers of physics with cold atoms.

The use of ultracold atoms to simulate relativistic as well as non-relativistic theories
and study the effect of disorder is an exciting field of research.
In a two-dimensional lattice subject to a non-Abelian gauge field,
one can induce not only localization transitions,
but also a transition from relativistic to non-relativistic theory
by tuning the lattice anisotropy.  A well known feature of the Dirac Hamiltonian 
is an extra term in the conductivity attributed to
{\it Zitterbewegung} (ZB)~\cite{ZB} corresponding to inter-band transitions.
It has been suggested that such a term is responsible for the finite conductivity of graphene
described by a massless Dirac energy spectrum~\cite{ZB}. In other words, it is ZB that
makes it impossible to localize relativistic particles, as it is
connected with the uncertainty of the position of a relativistic quantum particle
due to the creation of particle-antiparticle pairs.
Therefore, the origin of delocalization characterizing the non-Abelian system
that persists even for infinite disorder 
($\Lambda \rightarrow \infty$) can be attributed to ZB.

The detection of relativistic particle and a transition from non-relativistic to relativistic dispersion
in cold atoms in optical lattices was recently discussed; it was shown that the relativistic dispersion can be detected using atomic density profiles as well as Bragg spectroscopy ~\cite{duan}.

Our detailed study for various values of $\alpha_2$ captures some of the universal features of non-Abelian systems. 
Exploration of the two-dimensional space ($\alpha_1$, $\alpha_2$) may reveal additional phenomena, and the richness of $U(N)$ gauge systems with $N > 2$ remains to be explored.  Moreover, the effects of interparticle interactions remain to be investigated \cite{gpe}.

\section{Acknowledgment}
We are grateful to Ashwin Rastogi for his efforts in initiating the study of the
continuum limit of the problem and Jay Hanssen for helpful discussions regarding experimental aspects of this work.
We would also like to thank Ian Spielman and Nathan Goldman for their comments and suggestions on the paper.

%\renewcommand{\theequation}{B-\arabic{equation}}
% redefine the command that creates the equation no.
%\setcounter{equation}{0}  % reset counter
%\section*{Appendix A: Non-Abelian Fields}
%\section*{Appendix B: Aharanov-Bohm-Casher Effect}


\begin{references}
\bibitem{mono}J. Ruseckas, G. Juzeliunas, P. Ohberg and M. Fleischhauer, Phys. Rev. Lett., {\bf 95} 010404 (2005).
\bibitem{NJP}D. Jaksch and P. Zoller, New. J. Phys. \textbf{5}, 56 (2003).
\bibitem{NA}K. Osterloh, M. Baig, L. Santos, P. Zoller and M. Lewenstein, Phys. Rev. Lett., {\bf 95}, 010403 (2005).
%\bibitem{Muller}E.  Mueller, Phys. Rev. A, {\bf 70}, 041603 (2004).
\bibitem{Escher} Erich J. Mueller, Phys. Rev. A, {\bf 70}, 041603(R) (2004).
\bibitem{chern} N. Goldman and P. Gaspard, Europhys. Lett., {\bf 78}, 60001, (2007).
%\bibitem{baer}M. Baer, "Beyond Born-Oppenheimer," Wiley, 2006
\bibitem{Prange} See for example, Richard Prange, ``The Quantum Hall Effect," Springer-Verlag, edited by Richard Prange and Steven Girvin, (1990).
%\bibitem{AC} Naomichi Hatano, R. Shirasaki and H. Nakamura, quant-ph/0701076v1
\bibitem{Ons} L. Onsanger, Phil. Mag, {\bf 43}, 1006, (1952).
%\bibitem{Harper} For a review, see J. B. Sokoloff, Phys. Rep. \textbf{126}, 189 (1985).
\bibitem{Harper} P. G. Harper, Phys Soc. A, {\bf 68}, 879, (1955); S. Aubry and G. Andre, Ann. Isr. Phys. Soc., {\bf 3}, 133 (1980).
\bibitem{Hofs} D. R. Hofstadter, Phys. Rev. B, {\bf 14}, 2239 (1976).
\bibitem{Drese}K. Drese and M. Holthaus, Phys. Rev. Lett., \textbf{78}, 2932, (1997).
\bibitem{Sok}For a review, see J. B. Sokoloff, Phys. Rep. {\bf 126}, 189 (1985).
\bibitem{prl} Indubala I. Satija, Daniel C. Dakin and Charles W. Clark, Phys. Rev. Lett.,
{\bf 97}, 21640 (2006), erratum at {\bf 98}, 269904 (2007).
\bibitem{duan} Shi-Liang Zhu, Baigeng Wang and L.M. Duan, Phys. Rev. Lett., {\bf 98} 260402 (2007).
\bibitem{Fradkin} L. Balents and M. Fisher, Phys. Rev. B, {\bf 56}, 12970 (1997).
\bibitem{Amin} Amin Jazeri and Indubala I. Satija, Phys. Rev. E, {\bf 63}, 036222 (2001).
%\bibitem{4level}R. Unanyan, B. Shore and K. Bergmann, Phys Rev A, {\bf 59}, 2910, (1999).
\bibitem{ZB}  M. I. Katsnelson, Eur. Phys. J. B, {\bf 51}, 157 (2006).
\bibitem{data} D. A. Steck, data available at {\it http://steck.us/alkalidata}.
\bibitem{RPCW} A. M. Rey, G. Pupillo, C. W. Clark and C. J. Williams, Phys. Rev. A {\bf 72}, 033616, (2005)
\bibitem{gpe} N. Goldman, Europhys. Lett. {\bf 80}, 20001 (2007).
\end{references}
\end{document}